\documentclass[prl,twocolumn,showpacs,preprintnumbers,amsmath,amssymb, superscriptaddress]{revtex4}
\usepackage{graphicx}
\usepackage{dcolumn,color}
\usepackage{bm}
\usepackage[percent]{overpic}
\usepackage{multirow}
\usepackage{epsfig}
\usepackage[english]{babel}

\begin{document}

\bibliographystyle{try}
\topmargin 0.1cm

\newcounter{univ_counter}
\setcounter{univ_counter} {0} \addtocounter{univ_counter} {1}
 \edef\HISKP{$^{\arabic{univ_counter}}$ }\addtocounter{univ_counter}{1}
 \edef\PI{$^{\arabic{univ_counter}}$ } \addtocounter{univ_counter}{1}
 \edef\GIESSEN{$^{\arabic{univ_counter}}$ } \addtocounter{univ_counter}{1}
 \edef\BASEL{$^{\arabic{univ_counter}}$ } \addtocounter{univ_counter}{1}
 \edef\GATCHINA{$^{\arabic{univ_counter}}$ }\addtocounter{univ_counter}{1}
 \edef\BOCHUM{$^{\arabic{univ_counter}}$ } \addtocounter{univ_counter}{1}
 \edef\FSU{$^{\arabic{univ_counter}}$ } \addtocounter{univ_counter}{1}

\title{First measurement of the helicity asymmetry for $\gamma p\rightarrow p\pi^0$ in the resonance region}

\affiliation{Helmholtz--Institut f\"ur Strahlen-- und Kernphysik, Universit\"at Bonn, Germany}
\affiliation{Physikalisches Institut, Universit\"at Bonn, Germany}
\affiliation{II.~Physikalisches Institut, Universit\"at Gie{\ss}en, Germany}
\affiliation{Physikalisches Institut, Universit\"at Basel, Switzerland}
\affiliation{Petersburg Nuclear Physics Institute, Gatchina, Russia}
\affiliation{Institut f\"ur Experimentalphysik I, Ruhr--Universit\"at Bochum, Germany}
\affiliation{Department of Physics, Florida State University, Tallahassee, FL 32306, USA}

\author{M.\hspace{0.5mm}Gottschall\hspace{0.5mm}\HISKP}
\author{A.V.\hspace{0.5mm}Anisovich\hspace{0.5mm}\mbox{\HISKP\hspace{-1.2mm}$^,$\GATCHINA}}
\author{B.\hspace{0.5mm}Bantes\hspace{0.5mm}\PI}
\author{D.\hspace{0.5mm}Bayadilov\hspace{0.5mm}\mbox{\HISKP\hspace{-1.2mm}$^,$\GATCHINA}}
\author{R.\hspace{0.5mm}Beck\hspace{0.5mm}\HISKP}
\author{M.\hspace{0.5mm}Bichow\hspace{0.5mm}\BOCHUM}
\author{S.\hspace{0.5mm}B\"ose\hspace{0.5mm}\HISKP}
\author{K.-Th.\hspace{0.5mm}Brinkmann\hspace{0.5mm}\HISKP\hspace{-1.2mm}$^,$\GIESSEN}
\author{Th.\hspace{0.5mm}Challand\hspace{0.5mm}\BASEL}
\author{V.\hspace{0.5mm}Crede\hspace{0.5mm}\FSU}
\author{F.\hspace{0.5mm}Dietz\hspace{0.5mm}\GIESSEN}
\author{H.\hspace{0.5mm}Dutz\hspace{0.5mm}\PI}
\author{H.\hspace{0.5mm}Eberhardt\hspace{0.5mm}\PI}
\author{D.\hspace{0.5mm}Elsner\hspace{0.5mm}\PI}
\author{R.\hspace{0.5mm}Ewald\hspace{0.5mm}\PI}
\author{K.\hspace{0.5mm}Fornet-Ponse\hspace{0.5mm}\PI}
\author{St.\hspace{0.5mm}Friedrich\hspace{0.5mm}\GIESSEN}
\author{F.\hspace{0.5mm}Frommberger\hspace{0.5mm}\PI}
\author{Ch.\hspace{0.5mm}Funke\hspace{0.5mm}\HISKP}
\author{A.\hspace{0.5mm}Gridnev\hspace{0.5mm}\GATCHINA}
\author{M.\hspace{0.5mm}Gr\"uner\hspace{0.5mm}\HISKP}
\author{E.\hspace{0.5mm}Gutz\hspace{0.5mm}\HISKP\hspace{-1.2mm}$^,$\GIESSEN}
\author{Ch.\hspace{0.5mm}Hammann\hspace{0.5mm}\HISKP}
\author{J.\hspace{0.5mm}Hannappel\hspace{0.5mm}\PI}
\author{J.\hspace{0.5mm}Hartmann\hspace{0.5mm}\HISKP}
\author{W.\hspace{0.5mm}Hillert\hspace{0.5mm}\PI}
\author{Ph.\hspace{0.5mm}Hoffmeister\hspace{0.5mm}\HISKP}
\author{Ch.\hspace{0.5mm}Honisch\hspace{0.5mm}\HISKP}
\author{I.\hspace{0.5mm}Jaegle\hspace{0.5mm}\BASEL}
\author{D.\hspace{0.5mm}Kaiser\hspace{0.5mm}\HISKP}
\author{H.\hspace{0.5mm}Kalinowsky\hspace{0.5mm}\HISKP}
\author{S.\hspace{0.5mm}Kammer\hspace{0.5mm}\PI}
\author{I.\hspace{0.5mm}Keshelashvili\hspace{0.5mm}\BASEL}
\author{F.\hspace{0.5mm}Klein\hspace{0.5mm}\PI}
\author{E.\hspace{0.5mm}Klempt\hspace{0.5mm}\HISKP}
\author{K.\hspace{0.5mm}Koop\hspace{0.5mm}\HISKP}
\author{B.\hspace{0.5mm}Krusche\hspace{0.5mm}\BASEL}
\author{M.\hspace{0.5mm}Kube\hspace{0.5mm}\HISKP}
\author{M.\hspace{0.5mm}Lang\hspace{0.5mm}\HISKP}
\author{I.\hspace{0.5mm}Lopatin\hspace{0.5mm}\GATCHINA}
\author{Y.\hspace{0.5mm}Maghrbi\hspace{0.5mm}\BASEL}
\author{K.\hspace{0.5mm}Makonyi\hspace{0.5mm}\GIESSEN}
\author{V.\hspace{0.5mm}Metag\hspace{0.5mm}\GIESSEN}
\author{W.\hspace{0.5mm}Meyer\hspace{0.5mm}\BOCHUM}
\author{J.\hspace{0.5mm}M\"uller\hspace{0.5mm}\HISKP}
\author{M.\hspace{0.5mm}Nanova\hspace{0.5mm}\GIESSEN}
\author{V.\hspace{0.5mm}Nikonov\hspace{0.5mm}\HISKP\hspace{-1.2mm}$^,$\GATCHINA}
\author{R.\hspace{0.5mm}Novotny\hspace{0.5mm}\GIESSEN}
\author{D.\hspace{0.5mm}Piontek\hspace{0.5mm}\HISKP}
\author{G.\hspace{0.5mm}Reicherz\hspace{0.5mm}\BOCHUM}
\author{T.\hspace{0.5mm}Rostomyan\hspace{0.5mm}\HISKP}
\author{A.\hspace{0.5mm}Sarantsev\hspace{0.5mm}\mbox{\HISKP\hspace{-1.2mm}$^,$\GATCHINA}}
\author{St.\hspace{0.5mm}Schaepe\hspace{0.5mm}\HISKP}
\author{Ch.\hspace{0.5mm}Schmidt\hspace{0.5mm}\HISKP}
\author{H.\hspace{0.5mm}Schmieden\hspace{0.5mm}\PI}
\author{R.\hspace{0.5mm}Schmitz\hspace{0.5mm}\HISKP}
\author{T.\hspace{0.5mm}Seifen\hspace{0.5mm}\HISKP}
\author{V.\hspace{0.5mm}Sokhoyan\hspace{0.5mm}\HISKP}
\author{A.\hspace{0.5mm}Thiel\hspace{0.5mm}\HISKP}
\author{U.\hspace{0.5mm}Thoma\hspace{0.5mm}\HISKP}
\author{M.\hspace{0.5mm}Urban\hspace{0.5mm}\HISKP}
\author{H.\hspace{0.5mm}van\hspace{0.5mm}Pee\hspace{0.5mm}\HISKP}
\author{D.\hspace{0.5mm}Walther\hspace{0.5mm}\HISKP}
\author{Ch.\hspace{0.5mm}Wendel\hspace{0.5mm}\HISKP}
\author{U.\hspace{0.5mm}Wiedner\hspace{0.5mm}\BOCHUM}
\author{A.\hspace{0.5mm}Wilson\hspace{0.5mm}\FSU}
\author{A.\hspace{0.5mm}Winnebeck\hspace{0.5mm}\HISKP\hspace{-1.5mm}.}

\collaboration{The CBELSA/TAPS Collaboration}\noaffiliation

\date{\today}
\begin{abstract}
The first measurement of the helicity dependence of the
photoproduction cross section of single neutral pions off protons is
reported for photon energies from 600 to 2300\,MeV, covering nearly
the full solid angle. The data are compared to predictions from the
SAID, MAID, and BnGa partial wave analyses. Strikingly large
differences between data and predictions are observed which are
traced to differences in the helicity amplitudes of well known and
established resonances. Precise values for the helicity amplitudes
of several resonances are reported.
\end{abstract}

%----------end of abstract-------------

\maketitle

The spin structure of the proton has been of topical interest since
the discovery that the quark spins constitute an unexpectedly small
fraction of the proton spin~\cite{CERN-EP-87-230}. Deep inelastic
scattering experiments have revealed gluonic contributions to the
proton spin to be consistent with zero, at least within the
admittedly large errors \cite{Franco:2012mi}. At low energies, a
different aspect of the proton spin structure is tested by a
comparison of the helicity dependence of the total $\gamma p$ cross
section integrated over all photon energies, $\int_0^\infty \mathrm
dE_\gamma\,(\sigma_{3/2} -\sigma_{1/2})/E_\gamma$, with the proton
magnetic moment~\cite{arXiv:nucl-ex/0603021}, a relation which is
known as Gerasimov-Drell-Hearn (GDH) sum rule
\cite{48796,SLAC-PUB-0187}. The subscripts denote the total
helicity, $h$\,=\,$1/2$ for photon and proton spin anti-aligned,
$h$\,=\,$3/2$ for both spins aligned. The GDH integral sums over all
energies and all final states. A breakdown of the GDH integral into
exclusive final states can provide a link between inclusive
properties of the proton like its magnetic moment and the
contributions of specific nucleon resonances to the GDH integral.

In this letter, we report the first measurement of the double
polarization observable
\begin{equation}
E=
\frac{\left(\frac{\mathrm d\sigma}{\mathrm d\Omega}\right)_{1/2}-
\left(\frac{\mathrm d\sigma}{\mathrm d\Omega}\right)_{3/2}}
{\left(\frac{\mathrm d\sigma}{\mathrm d\Omega}\right)_{1/2}+
\left(\frac{\mathrm d\sigma}{\mathrm d\Omega}\right)_{3/2}}
\label{E}
\end{equation}
for the exclusive reaction $\gamma p\to p\pi^0$ in the energy range
from 600 to 2300\,MeV, and compare the results with predictions of
well established partial wave analyses (PWA) like
SAID~\cite{SAID-SN11,SAID-CM12}, MAID~\cite{Drechsel:2007if}, and
BnGa~\cite{Anisovich:2011fc}. The double polarization observable $G$
\cite{Thiel:2012yj}  -- governing the correlation between linearly
polarized photons and longitudinally polarized target protons --
revealed remarkable differences in the predictions of the three 
partial wave analysis groups even in the well-studied 2$^{\rm nd}$
resonance region around $E_\gamma=$750\,MeV~\cite{Comment}. 
Here, we extend the covered energy regime to 2.3\,GeV for the 
double polarization observable $E$. So
far, data on $E$ were published up to 780\,MeV for a limited angular
range~\cite{Ahrens:2002,Ahrens:2004}; further data from CLAS exist and were reported at a
conference~\cite{Iwamoto:2012zza}.

%------------Figure 1---------------------------------------
\begin{figure*}[pt]
 \begin{tabular}{lcr}
  \includegraphics[width=0.485\textwidth]{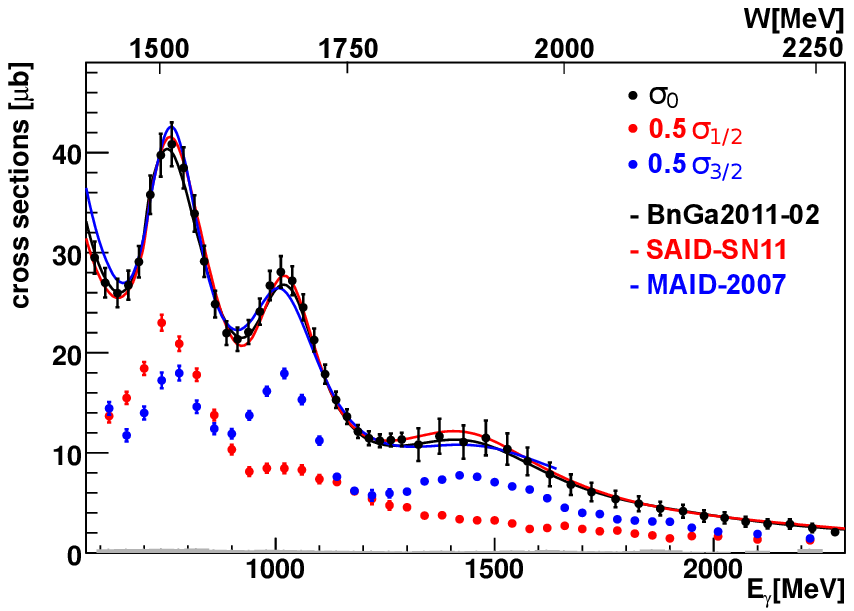}
&~~~&
  \includegraphics[width=0.489\textwidth]{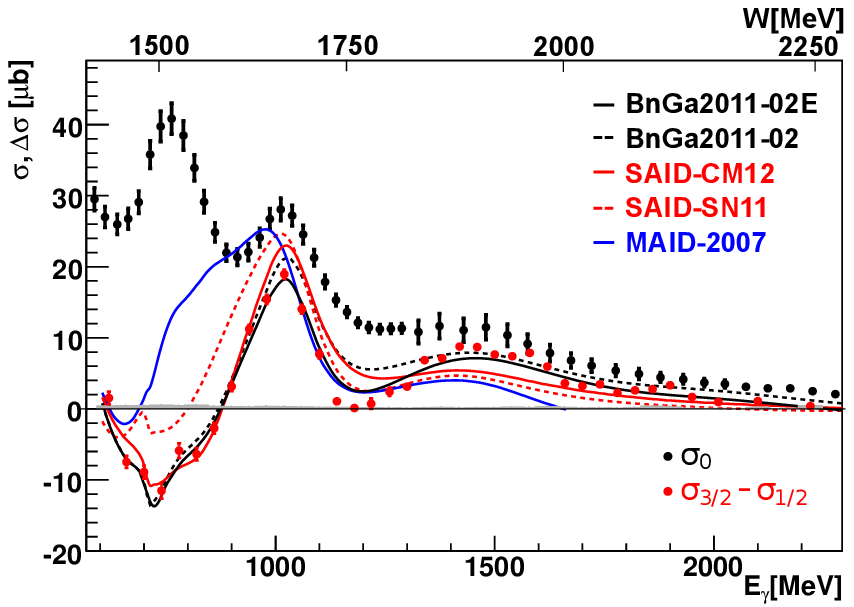}
 \end{tabular}
\caption{\label{pic:Sigma}The total cross section
$\sigma_0$~\cite{Pee:2007} plotted together with  its two helicity
components (left) and $\sigma_{3/2}$\,-$\,\sigma_{1/2}$ (right) as a
function of $E_{\gamma}$. The error bars give the statistical
errors, the systematic errors are shown as a dark gray band.}
\end{figure*}
%---------------------------------------------------

The experiment was carried out using the tagged photon beam of the
ELectron Stretcher Accelerator ELSA at Bonn \cite{Hillert:2006yb}.
Photons with circular polarization~\cite{Olsen:1959zz},
\begin{equation}
P_{\odot}=\frac{4x-x^2}{4-4x+3x^2}P_{\rm e^-}\quad\mbox{with~}x=\frac{E_{\gamma}}{E_{\rm e^-}},
\end{equation}
were produced by scattering a 2.4\,GeV beam of longitudinally
polarized electrons with polarization $P_{\rm e^-}$ off a
bremsstrahlung target. The electrons were deflected by a magnet into
a tagging hodoscope which defines the energy of the bremsstrahlung
photons. The electron polarization ($P_{\rm e^-}$\,$\approx$\,0.60)
was monitored in a M{\o}ller polarimeter \cite{kammer}. The photon
beam impinged on a butanol ($\rm C_4H_9OH$) target, which was
polarized by dynamic nucleon polarization \cite{Dutz:2004eb}. An
average polarization of ${P_T}$\,$\approx$\,0.71 was obtained.

Particles produced by photoproduction off the target were detected
in a 4$\pi$ geometry using several sub-detectors. The target was
surrounded by an inner scintillating fiber detector designed to
detect charged particles \cite{Suft:2005cq}, and by the CsI(Tl)
Crystal Barrel detector \cite{CB}. A forward calorimeter of 90
CsI(Tl) crystals and a TAPS wall of 216 BaF$_2$ crystals \cite{TAPS}
completed the calorimeter setup in the forward direction. For
charged particle detection, both forward calorimeters were equipped
with plastic scintillation counters in front of the crystals. With
this setup, the four-momenta of neutral mesons decaying into photons
could be determined by measuring their energies and directions from
the target center to the shower center. The reaction $\gamma p\to
p\pi^0\to p\gamma\gamma$ was reconstructed by a series of kinematic
cuts. Events with exactly two photons and a proton were used. The
latter was demanded in order to suppress background from reactions
off neutrons. For the proton candidate, only a hit in one of the
charged particle detectors was required to include also low energy
protons, which did not reach the calorimeter. The invariant mass of
the two photons and the missing mass was calculated and a
{$\pm2\sigma$} cut around the $\pi^0$ and the proton mass was
applied. The azimuthal angle between the direction of proton and
meson was asked to be 180$^{\circ}$ within a $\pm2\sigma$ window
(coplanarity). To remove untagged events originating from photons
below the tagging threshold (due to random coincidences), the beam
photon energy was also calculated from the reaction and compared to
the measured photon energy. Finally, a time coincidence was required
between the tagger hit and the reaction products and random time
background was subtracted. 

The helicity-dependent differential cross section for
photoproduction of neutral pions off longitudinally polarized
protons can be written as
\begin{equation}
\frac{\mathrm d\sigma}{\mathrm d\Omega} =
\frac{\mathrm d\sigma_0}{\mathrm d\Omega}\,(1\pm P_TP_{\odot}{E})\,.
\label{sigma}
\end{equation}
When a butanol ($\rm C_4H_9OH$) target is used, unpolarized $(u)$
free protons $(f)$ as well as nucleons bound $(b)$ in carbon or
oxygen contribute to the count rate, in addition to the polarized
$(p)$ free protons. For the same number of beam photons for both
photon helicities, and under the assumption that (unpolarized)
nucleons bound in carbon have the same response to impinging photons
as nucleons bound in oxygen, we get for beam and target polarized in
the same direction the yield $N_{3/2}$\,=\,
$N^{f,p}_{3/2}$\,+\,$N^{f,u}$\,+\,$N^b$ and in opposite direction
$N_{1/2}$\,=\,$N^{f,p}_{1/2}$\,+\,$N^{f,u}$\,+\,$N^b$, which leads
to {\begin{equation} {E}=
\frac{N_{1/2}-N_{3/2}}{N_{1/2}+N_{3/2}}\cdot \frac{1}{d}\cdot
\frac{1}{P_{\odot}P_T}\,.
\label{counts}
\end{equation}}
The bound nucleons are taken into account by the dilution factor
$d=\frac{N^f}{N^f+N^b}$. It was determined by an additional
measurement using a carbon foam target with approximately the same
density, size, and environment as the carbon and oxygen part of the
butanol target. The fraction of carbon (and oxygen) in butanol was
then determined using two alternative methods \cite{Diss}: by
fitting a) the missing mass and b) the coplanarity distribution
outside of the signal of the free protons. The resulting scaling
factor $s(E_{\gamma})$ for each energy bin was then used to
determine {\begin{equation} d(E_{\gamma},\theta_{\pi})= \frac{N_{\rm
C_4H_9OH}(E_{\gamma},\theta_{\pi})-s(E_{\gamma})\cdot
N_{C}(E_{\gamma},\theta_{\pi})}{N_{\rm
C_4H_9OH}(E_{\gamma},\theta_{\pi})}\,.
\end{equation}}
Fig.~\ref{pic:Sigma} shows on the left the total cross section
$\sigma_0$ and its decomposition into the two helicity components
$\sigma_0$\,=\,$1/2 (\sigma_{1/2}$\,+\,$\sigma_{3/2}$). The latter
cross sections were calculated from the values for $E$ reported
below - and extrapolated into the unmeasured angular bins by the
BnGa-PWA - and the total cross section $\sigma_0$ (from the
BnGa-PWA). The figure reveals important details of the spin
structure of the photo-excitation of the proton; $\sigma_{1/2}$ and
$\sigma_{3/2}$ evidence quite different structures. Both cross
sections show a peak at $W$\,$\approx$\,1500\,MeV in $\sigma_{1/2}$
and, slightly shifted, at $W$\,$\approx$\,1520\,MeV in
$\sigma_{3/2}$. The latter peak can be assigned to the
N(1520)$3/2^-$ and its large helicity coupling to $h$\,=\,$3/2$. The
lower mass peak cannot be solely due to the N(1535)$1/2^-$:
contributions from other resonances or from background amplitudes,
e.g. in the $1/2^-$ partial wave, must be significant. The
N(1680)$5/2^+$ contributes nearly only to $\sigma_{3/2}$, where a
clear peaking structure is observed. A broad structure at
$W$\,$\approx$\,1900\,MeV is very visible in the $\sigma_{3/2}$
excitation curve and seems to be absent in $\sigma_{1/2}$. Photons
in the $E_{\gamma}\approx$\,1500\,MeV range may excite some of the
resonances with spin $J=1/2,\cdots,7/2$ in the
$W$\,=\,1900\,-\,2000\,MeV mass region. Their contribution is also
well visible in the cross section difference
$\sigma_{3/2}$\,-\,$\sigma_{1/2}$\,=$\,-E\cdot 2\sigma$ in
Fig.~\ref{pic:Sigma} on the right. Obviously, they are
preferentially excited by the $A_{3/2}$ helicity amplitude.

\begin{figure}[pt]
\includegraphics[width=0.46\textwidth,height=0.495\textheight]{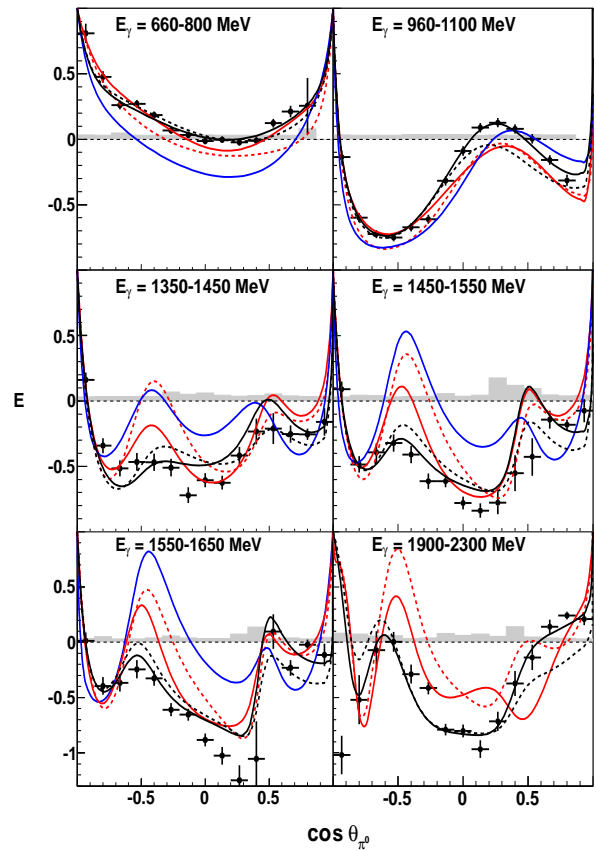}
\caption{\label{pic:results}The helicity asymmetry $E$ as a function
of $\cos\theta_{\pi}$ for selected $E_{\gamma}$ bins. PWA
predictions: black dashed curve: BnGa2011-02, blue solid curve:
MAID, red solid curve: CM12, red dashed curve: SN11. Fit to the data
points (BnGa2011-02E): black solid curve.}
\end{figure}

The helicity asymmetry $E$ in eq.(\ref{counts}) is a function of the
pion production angle $\theta_{\pi}$. Fig.~\ref{pic:results} shows
selected results. There are very significant changes in these
distributions as a function of energy. Obviously, the quantity $E$
is  very sensitive to the contributions from baryon resonances.
The lowest incident photon energy bin covers the part of the region
with positive $E$. The shape indicates a strong contribution from
the $J^P=1/2^-$ and $J^P=3/2^-$ partial waves, with 
N(1535)$1/2^-$ and N(1520)$3/2^-$ being the dominant resonances
in this energy range. The asymmetry in the angular distribution
reflects the presence of several weakly contributing resonances,
like N(1440)$1/2^+$ and $\Delta(1232)3/2^+$.
The next bin, $E_{\gamma}$\,=\,960\,-\,1100\,MeV, covers 
N(1680)$5/2^+$. Together with the $J^P=1/2^-$ partial wave, it
produces a W-shaped angular distribution. The strong
forward-backward asymmetry signals contributions from other partial
waves, among which the $D_{33}$ partial wave including the
$\Delta(1700)3/2^-$ plays an important role.
We have split the 4$^{\rm th}$ resonance region from 1350 to
1650\,MeV into three 100\,MeV slices. The basic structure remains
over the full energy range, which suggests a sizeable contribution
from a $J^P=7/2^+$ resonance, which would lead to a three minima
structure in the angular distribution. There are only relatively
small differences when the three energy bins are compared: 
$\Delta(1950)7/2^+$ seems to be the dominant contribution. The
highest mass bin exhibits strong structures but no clear pattern;
several resonances seem to make comparable contributions to the
data.
\begin{table*}[!ht]
\begin{scriptsize}   \begin{center}\renewcommand{\arraystretch}{1.3}
    \begin{tabular}{|c|ccccc|c|ccccc|}
     \hline\hline
 & MAID2007 &\hspace{3mm}SN11\hspace{3mm} &\hspace{3mm}CM12\hspace{3mm}  & BnGa2011&BnGa2011E &  ~   & MAID2007 & \hspace{3mm}SN11\hspace{3mm} & \hspace{3mm}CM12\hspace{3mm} & BnGa2011&BnGa2011E   \\
     \hline
   N(1440) &&&&&& N(1520) &&&&&\\\hline
     M            & 1440    & 1485     &1485      & 1430$\pm$8 & 1430$\pm$8 && 1530  & 1515      & 1515$\pm$3 & 1517$\pm$3 & 1516$\pm$2 \\
     $\Gamma$     & 350     & 284      &284       & 365$\pm$35 & 360$\pm$30 && 130   & 104       & 104$\pm$5  & 114$\pm$5  & 113$\pm$5    \\
     BR(N$\pi$)   &  70     & 79       &79        & 62$\pm$3   & 62$\pm$3   && 60    & 63        & 63$\pm$3   & 62$\pm$3   &  62$\pm$2  \\
     A$_{1/2}$    & -61     &-58$\pm$1 &-56$\pm$1 & -61$\pm$6  & -62$\pm$8  && -27   & -16$\pm$2 & -19$\pm$2  & -22$\pm$4  & -20$\pm$3  \\
     A$_{3/2}$    & ~~      &~         &~~        &  ~~        &        && 161   & 156$\pm$2 & 141$\pm$2  & 131$\pm$10 & 131$\pm$7 \\
     \hline
     N(1535) &&&&&& N(1650)   &&&&&   \\\hline
     M          &  1535 & 1547   &1547     &1519$\pm$5 &1518$\pm$4&&  1690       & 1635     &1635     &1651$\pm$6&1651$\pm$6\\
     $\Gamma$   & 100  & 188    &188      &128$\pm$14 &125$\pm$10 &&   100       & 115      &115      &104$\pm$10&102$\pm$10 \\
     BR(N$\pi$) & 40   & 36     &36       &54$\pm$5   & 55$\pm$5  &&   85         & 100      &100      &51$\pm$4  &50$\pm$4 \\
     A$_{1/2}$  & 66   &99$\pm$2&128$\pm$4&105$\pm$10 & 105$\pm$9  && 33       & 65$\pm$25&55$\pm$30&33$\pm$7  &33$\pm$7\\
     \hline
     $\Delta$(1620) &&&&&& $\Delta$(1950)   &&&&&   \\
     \hline
     M      & 1620 & 1615   & 1615   &1600$\pm$8 & 1598$\pm$6     &&   1945 & 1921    & 1921   &1915$\pm$6 & 1915$\pm$5\\
     $\Gamma$   & 150  & 147    &  147   &130$\pm$11 & 130$\pm$8 &&   280  & 271      &  271   &246$\pm$10 &249$\pm$8     \\
     BR(N$\pi$) & 25   & 32     &  32    & 28$\pm$3  & 28$\pm$3  &&   40   & 47       &  47    & 45$\pm$2  & 46$\pm$2    \\
     A$_{1/2}$  & 66   &64$\pm$2&29$\pm$3& 52$\pm$5  & 52$\pm$5 && -94   &-71$\pm$2&-83$\pm$4& -71$\pm$4  & -70$\pm$5  \\
     A$_{3/2}$  &        &&&  &&& -121   &-92$\pm$2&-96$\pm$4& -94$\pm$5  & -93$\pm$5   \\  \hline\hline
    \end{tabular}
   \end{center}
\end{scriptsize}
\caption{\label{tab:nstars}Properties of selected low-mass nucleon
resonances. Breit-Wigner mass, width, and $N\pi$ partial decay width as defined in~\cite{Anisovich:2011fc}
are given in MeV, the helicity amplitudes in $10^{-3}$\,GeV$^{-1/2}$. }
\end{table*}

Next we compare the data in Figs.~\ref{pic:Sigma} and
\ref{pic:results} with predictions from SAID (SN11 and CM12
\cite{SAID-SN11, SAID-CM12}), MAID \cite{Drechsel:2007if} and
BnGa2011-02 \cite{Anisovich:2011fc}. In addition, a fit to the data
within the BnGa-PWA (BnGa2011-02E) is shown. First we note that all
PWAs give a reasonable description of $\sigma^{\gamma p\to p\pi^0}$.
But we find large and unexpected discrepancies already at rather low
energies, in the region of the four-star resonances N(1440)$1/2^+$,
N(1535)$1/2^-$ and N(1520)$3/2^-$. In the $\Delta(1950)7/2^+$
region, MAID and SAID show a striking enhancement at
$\cos\theta_{\pi}\approx -0.45$. In the MAID online version, the
enhancement disappears when the helicity couplings of the
N(1520)$3/2^-$ are reduced by a factor of 4. The N(1520)$3/2^-$,
however, is the only resonance in this partial wave; hence, such a
large effect, which is presumably due to interference with other
partial waves, is unexpected. In SN11, this structure is smaller
than in MAID, and is further reduced in CM12. The origin of the
structure, also visible in the SAID prediction, remains unclear.
None of the fits reproduces the data over the full angular range.
However, BnGa2011-02 required only a slight adjustment of free
parameters to get close to the data points (solution BnGa2011-02E).
For examples see Table~\ref{tab:nstars}. It compares selected
results on low-mass excitations of the nucleon for MAID and for the
two latest solutions from SAID and BnGa, respectively. One formal
difference is remarkable between MAID and SAID on the one hand, and
BnGa on the other hand: masses, widths and branching ratios are
fixed from $\pi$N elastic scattering in MAID and SAID. This approach
excludes the possibility of photoproduction providing more than the
photo-couplings of resonances. Results of the multichannel BnGa-PWA
demonstrate however the importance of including information from
photon induced and especially double polarization measurements for
the determination of nucleon resonance parameters in general.

Finally, we come back to the GDH integral. The contribution of
single $\pi^0$ production to the GDH integral is evaluated. The
total GDH integral up to 2.9\,GeV  was determined to $I^{\rm
2.9}_{\text{tot}}=226\pm5^{\text{stat}}\pm12^{\text{syst}}$\,$\mu$b
\cite{Ahrens:2000bc,Ahrens:2001qt,Dutz:2003mm,Dutz:2004zz,Dutz:2005ns};
the results were summarized in \cite{arXiv:nucl-ex/0603021}. Here, a
contribution of $27.5$\,$\mu$b had been subtracted to account for
the description of the very low energy region by MAID
\cite{Drechsel:2007if}. Restricted to a maximum energy of 2.3\,GeV,
the total integral results in $227\pm5^{\text{stat}}$\,$\mu$b and we
find $159\pm8^{\text{stat}}$\,$\mu$b for the contribution of single
$\pi^0$ production. Up to 0.6\,GeV photon energy, this contribution
amounted to $156\pm8^{\text{stat}}$\,$\mu$b: above 
$\Delta(1232)$, there is practically no contribution of single
$\pi^0$ production to the GDH integral.

Summarizing, we have reported the first measurement of the helicity
dependent photoproduction cross section for photons at
$E_{\gamma}$\,=\,0.6-2.3\,GeV with nearly full solid angular
coverage. The observable $E$ is shown to be highly sensitive to the
contributions from $s$-channel resonances. Even after many years of
studying the simplest photoproduction reaction, $\gamma p\to
p\pi^0$, the new data reveal very significant discrepancies in
comparison with model predictions. As expected, it is obvious that a
fit to differential cross sections and to single polarization
observables alone is not sufficient to arrive at unambiguous
solutions.

We thank the technical staff of ELSA and the par\-ti\-ci\-pating
institutions for their invaluable contributions to the success of
the experiment. We acknowledge support from the \textit{Deutsche
Forschungsgemeinschaft} (SFB/TR16) and \textit{Schweizerischer
Nationalfonds}.

\clearpage
\end{document}